\documentclass[twocolumn,letter]{jpsj2} 
\def\d{{\rm d}}

\def\dl{{d \over{dl}} \,}

\def\e{{\rm e}}

\def\kf{k_{\rm F}}

\def\ggs{\buildrel\textstyle > \over {\hbox{\raise0.2ex\hbox{$\sim$}}}}
\def\lls{\buildrel\textstyle < \over {\hbox{\raise0.2ex\hbox{$\sim$}}}}
\def\gsim{\,\lower0.75ex\hbox{$\ggs$}\,}
\def\lsim{\,\lower0.75ex\hbox{$\lls$}\,}

\def\tt{\tilde{t}}
\def\im{{\rm i}}
\def\ie{{\it i.e.}, }

\def\et{{\it et al.}, }
\def\delx{\partial_x}

\def\ti #1{\tilde #1}
\def\e{{\rm e}}

\def\delx{\partial_x}
\def\al{\alpha}
\def\vr{\vec r}
\def\ti#1{\tilde #1}
\def\delx{\partial_x}
\def\dely{\partial_y} 
\def\jo #1#2#3#4{#1 {\bf #2} (#3) #4}   

\def\PRB{Phys.\ Rev.\ B}
\def\PRL{Phys.\ Rev.\ Lett.}

\def\JPSJ{J.\ Phys.\ Soc.\ Jpn.}

\def\runtitle
{Incommensurate Mott Insulator in One-Dimensional Electron Systems close
to quarter filling}
\def\runauthor{Hideo Yoshioka, Hitoshi Seo and
Hidetoshi Fukuyama}

\title
{Incommensurate Mott Insulator in One-Dimensional Electron Systems close
to Quarter Filling}

\author{Hideo \textsc{Yoshioka}$^1$\thanks{E-mail address: 
h-yoshi@cc.nara-wu.ac.jp}, Hitoshi \textsc{Seo}$^2$ and  
Hidetoshi \textsc{Fukuyama}$^3$} 

\inst{$^1$Department of Physics, Nara Women's University, Nara 630-8506
\\
$^2$
Non-Equilibrium Dynamics Project, ERATO-JST, 
c/o KEK, Tsukuba 305-0801\\ 
$^3$International Frontier Center for Advanced Materials (IFCAM), 
IMR, Tohoku University, Sendai 980-8577
}

\recdate{\today}

\abst{
A possibility of a metal-insulator transition in molecular conductors 
has been studied for systems composed of donor molecules and fully ionized anions 
with an incommensurate ratio close to 2:1 based on 
a one-dimensional extended Hubbard model, where the donor carriers are  
slightly deviated from quarter filling and  
under an incommensurate periodic potential from the anions. 
By use of the renormalization group method,
interplay between commensurability energy on the donor lattice and that from  
the anion potential has been studied and it has been found that  
an ``incommensurate Mott insulator" can be generated. 
This theoretical finding will explain the metal-insulator transition
observed in (MDT-TS)(AuI$_2$)$_{0.441}$.           
}

\kword{Mott insulator, incommensurate potential, organic conductors, metal-insulator transition}

\begin{document}
\maketitle

Most of the conducting molecular crystals are realized by 
combining two kinds of molecules, $A$ and $B$, 
with commensurate composition ratios. 
Typical examples are the well-studied 2:1 compounds, \ie $A_2 B$, 
which show a variety of phases 
such as Mott insulator, charge order, superconductivity 
and so on.~\cite{Ishiguro-Yamaji-Saito, ChemRev}  
In the compounds, the molecule $B$ is usually fully ionized either as $-1$ or $+1$ 
to form a closed shell, 
and as a consequence the energy band formed by HOMO or LUMO of $A$  
is quarter-filled as a whole in terms of holes or electrons, respectively. 

Recently, molecular conductors with incommensurate (IC) 
composition ratios close to 2:1 have been synthesized 
based on new donor molecules.~\cite{Takimiya, Kawamoto, Kawamoto-5} 
(MDT-TSF)$X_n$ and (MDT-ST)$X_n$ 
($X$ = I$_3$, AuI$_2$ or IBr$_2$, $n$ $\simeq$ 0.42 -- 0.45)
show metallic behavior 
and undergo superconducting transition at about $T_c$ = 4 K at ambient pressure.~\cite{Takimiya,Kawamoto} 
In contrast, in (MDT-TS)(AuI$_2$)$_{0.441}$ 
a metal-insulator (MI) crossover occurs 
where the temperature dependence of resistivity 
displays a minimum at $T_\rho$ = 85 K.~\cite{Kawamoto-5} 
In addition, an antiferromagnetic transition takes place 
at $T_{\rm N}$ = 50 K. 
When pressure is applied to this compound, $T_\rho$ decreases and 
the superconducting phase appears above $P_c$ = 10.5 kbar ($T_c$ = 3 K). 
All these compounds are isostructural with alternating donor and anion layers. 
Since the anions are fully ionized as 
$X^{-}$ as in the 2:1 compounds,~\cite{Takimiya}  
the electronic properties can be 
attributed to the donors with the IC band-filling 
slightly larger than 3/4 for the HOMO bands. 
The extended H$\ddot{\rm u}$ckel scheme predicts 
two-dimensional (2D) Fermi surfaces 
which are similar to each other.~\cite{Kawamoto,Kawamoto-5}  
It should be noticed that 
the anions in these compounds 
are not randomly distributed in the layers,
but are found by the X-ray scattering experiments to form regular IC lattices with a different periodicity
from the donors.~\cite{Takimiya,Kawamoto} 

The metallic state observed in these compounds 
is naturally expected from the IC band-filling 
since the system would avoid insulating states 
due to strong correlation such as Mott insulator or charge order. 
On the other hand, 
it is difficult to understand
the strong-coupling nature of 
the insulating ground state in (MDT-TS)(AuI$_2$)$_{0.441}$, 
deduced from $T_\rho \neq T_{\rm N}$, 
which is to be explored in this Letter; 
if it is the weak-coupling spin-density-wave state 
due to the nesting of the Fermi surface, 
$T_\rho = T_{\rm N}$ would be expected.~\cite{Ishiguro-Yamaji-Saito} 

\begin{figure}
 \centerline{\includegraphics[width=7.0truecm]{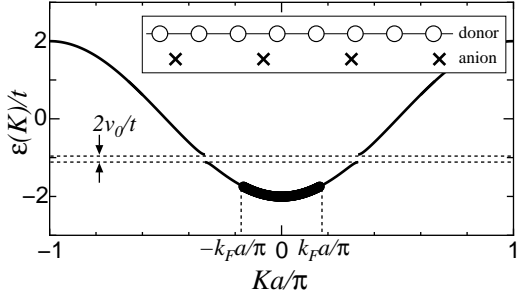}}
\vspace*{-1em}
\caption{Energy dispersion in the presence of the anion potential $v_0 = 0.1t$
where the occupied one-particle states are expressed by the thick curve.
The figure is written in the case of $n=0.33$ to clarify the
 characteristics of the present model.
Inset: a schematic representation of our model.
} 
\label{fig:band}
\end{figure}

We consider a one-dimensional (1D) model 
in order to capture the essence 
of the MI transition in a more controlled way than considering 
a 2D model relevant to experiments. 
Our 1D model consists of $N_L$ donor
molecules coupled with $N$ anions both forming regular lattices, 
as shown in the inset of Fig.~\ref{fig:band}. 
The donors are modeled by the 1D extended Hubbard model, known 
to be relevant for typical 2:1 systems,~\cite{Seo-Hotta-Fukuyama,Yoshioka} 
and the small potential from the anions is added, 
which is crucial for the insulating state to appear. 
The Hamiltonian is written as follows, 
\begin{eqnarray} 
 {\cal H} &=& -t \sum_{j,s}(c_{j+1,s}^\dagger c_{j,s} + h.c.) + 
\frac{U}{2} \sum_{j,s} n_{j,s} n_{j,-s} \nonumber \\
&+& V \sum_{j} n_j n_{j+1} + 
\sum_{j} v_j n_j,
\label{eqn:H} 
\end{eqnarray}
where $t$, $U$ and $V$ are respectively  
the transfer energy between the nearest-neighbor donor sites, 
the on-site repulsive interaction and the nearest-neighbor repulsion; 
the creation operator at the $j$-th site with spin $s$=$\pm$ is denoted as
$c^\dagger_{j,s}$, $n_{j,s} = c^\dagger_{j,s} c_{j,s}$ and 
$n_j = \sum_s n_{j,s}$. 
Since the fully ionized anions form a regular lattice, 
the anion potential at the $j$-th site, $v_j$, can be expressed as 
$v_j = N_L^{-1} \sum_{m=-\infty}^{\infty} v(mQ) \e^{\im m Q j a}$ 
where $Q = 4k_F$, $k_F = \pi n/(2a)$ is the Fermi wavenumber, $n=N/N_L$ is the carrier density
(we take the hole picture in the following) in the donor chain and 
$a$ is the spacing between donor sites.  
In the following, 
we consider only the relevant $\pm Q$ component of the potential,
$v(\pm Q) \equiv v_0 \e^{\pm \im \chi}$. 
This can lead to a gap, $2 v_0$, at $\pm 2 \kf$ in the non-interacting 
band, which we assume to be small compared to the band width. 
Then the system becomes effectively 
half-filled in reference to the IC lattices as is seen in Fig.1. 

We derive an effective Hamiltonian for low energy scale 
in terms of phase variables following the just 
quarter-filling case.\cite{Yoshioka} 
To the lowest order of the normalized anion potential 
$\delta \equiv v_0/\left\{ \epsilon(3k_F) - \epsilon(k_F) \right\}$ with $\epsilon(K) = -2t \cos
Ka$, 
the phase Hamiltonian is obtained as  
${\cal H}_{\rm eff} = {\cal H}_{\rho} + {\cal H}_{\sigma} + {\cal
H}_{\rm \rho \sigma}$. 
Here ${\cal H}_{\rho}$, ${\cal H}_{\sigma}$ and ${\cal H}_{\rho
\sigma}$ are respectively the charge part, the spin part and the term 
mixing both degrees of freedom. 
The spin part, ${\cal H}_{\sigma}$ has the same form as that of the
1D Hubbard model, so the spin excitation becomes gapless.~\cite{Emery}
The term ${\cal H}_{\rho \sigma}$ is expressed by the product of the
non-linear terms seen in ${\cal H}_\rho$ and ${\cal H}_\sigma$, 
and then has a larger scaling dimension.
Hence we can neglect it~\cite{noteTsuchiizu-Furusaki}. 
Therefore the properties 
of the charge degree of freedom are determined by ${\cal H}_{\rho}$, expressed as 
\begin{eqnarray}
{\cal H}_\rho &=& \frac{v_\rho}{4\pi} \int d x 
\left\{ \frac{1}{K_\rho} (\partial_x \theta_\rho)^2 + K_\rho (\partial_x \phi_\rho)^2 \right\} \nonumber \\ 
&+& \frac{g_{3 \bot}}{(\pi \alpha)^2} \int dx \cos (2 \theta_\rho + 3 \chi) \nonumber \\
&+& \frac{g''_{3 \bot}}{(\pi \alpha)^2} \int dx \cos (2 \theta_\rho + 5 \chi - q_0 x) \nonumber \\
&+& \frac{g_{1/4}}{2 (\pi \alpha)^2} \int dx \cos (4 \theta_\rho + 8 \chi - q_0 x). 
\label{eqn:phaseH}
\end{eqnarray}
Here $q_0 = 2 \pi/a - 8\kf = 2 \pi (1-2n)/a$ is the misfit parameter, 
$v = 2 ta \sin \kf a$, $v_\rho = v \sqrt{B_\rho A_\rho}$
and $ K_\rho = \sqrt{B_\rho/A_\rho}$ 
with $A_\rho = 1+ (g_{4||} + g_{4\bot} + g_{2||} + g_{2\bot} -
g_{1||})/(\pi v)$ and $B_\rho = 1 + (g_{4||} + g_{4\bot} - g_{2||} -
g_{2\bot} + g_{1||})/(\pi v)$. $\alpha^{-1}$ is the ultra-violet
cut-off ($\alpha \sim a$).  
The interaction parameters are written as 
\begin{eqnarray}
%
%
 g_{1||} &=& Va \cos 2 k_F a \nonumber \\ &-& 4 D_1 \left(Va \cos 2 k_F a\right)  \left(Va \cos 4 k_F a\right),  \\
 g_{2\bot} &=& \frac{Ua}{2} + Va - 2 D_1 \nonumber \\ 
& & \hspace*{-5em} \times   
\left\{
\left( \frac{Ua}{2}  + Va \cos 2 k_F a \right)^2 
 + \left( \frac{Ua}{2}  + Va \cos 4 k_F a \right)^2 
\right\},  \nonumber \\ && \\
 g_{2||} &=& Va \nonumber \\ &-& 2 D_1
\left\{
\left( Va \cos 2 k_F a \right)^2 + \left( Va \cos 4 k_F a \right)^2 
\right\}, \\
g_{3 \bot} &=& -4 \delta (\frac{Ua}{2} + Va \cos 2 k_F a ) \nonumber \\ 
&+& 4 \delta D_2 (\frac{Ua}{2} + Va \cos 4 k_F a) \nonumber \\ 
& & \times \left(Ua + Va \cos 2 k_F a + Va \cos 6k_F a \right), \\
%
g''_{3 \bot} &=& - 4 \delta D_2 (\frac{Ua}{2} + Va \cos 4 k_F a ) \nonumber \\ && \times (Ua + 2 Va \cos 2 k_F a ), \\
%
g_{4 \bot} &=& \frac{Ua}{2} + Va - 2 D_2 \left( \frac{Ua}{2}  + Va \cos 4 k_F a \right)^2, \\
g_{4 ||} &=&  Va  - 2 D_2 \left( Va \cos 4 k_F a \right)^2, \\
g_{1/4} &=& \frac{X}{2 (\pi \alpha)^2 (\epsilon(3k_F) - \epsilon(k_F))^2},  \\
 X &=& 2 \Big\{
(2 Va \cos 2 k_F a)^2 ( Ua + 2 Va \cos 4 k_F a) \nonumber \\
&+& (Ua + 2Va \cos 2 k_F a)^2 (Ua + 4Va \cos 4k_F a) \nonumber \\
&+& (2Va \cos 2k_F a) (Ua + 2Va \cos 2 k_F a) \nonumber \\ 
& & \times (Ua + Va \cos 2 k_F a + Va \cos 6 k_F a)
\Big\} \nonumber \\ 
&+& (Ua + 2Va \cos 2 k_F a)^2 \nonumber \\ & & \times (Va \cos 2 k_F a + Va \cos 6 k_F a), 
\end{eqnarray}
with 
\begin{eqnarray}
 D_1 &=& 
\frac{1}{4 \pi v(3k_F)} \nonumber \\ 
&& \times \ln \frac{\epsilon(3k_F) - \epsilon(k_F) + v(3k_F)k_F}{\epsilon(3k_F) - \epsilon(k_F)- v(3k_F)k_F}, \\ 
 D_2 &=& 
\frac{1}{4\pi} \frac{2 k_F}{\epsilon(3k_F) - \epsilon(k_F)},
\end{eqnarray}
where $v(3k_F) = 2ta \sin 3k_F a$. 

In eq.(2), there are three non-linear terms. 
First, the half-filling Umklapp term, $g_{3 \bot}$, is generated 
by the anion potential $\delta$ 
because the band is effectively half-filled as seen in Fig.1. 
This can lead to  a Mott insulator, 
as we will later show explicitly.  
We call the state as {\it IC Mott insulator} since it has 
a periodicity not matching with the donors but with the anions. 
However, 
it is not trivial whether this IC Mott insulator can be realized, 
and if so, in which condition it is stabilized, 
in contrast to the half-filled Hubbard model where 
infinitesimal on-site repulsion stabilizes the Mott insulator.~\cite{Emery}
This is because of the presence of the other two non-linear terms in eq.(2), 
the ``quarter-filling" Umklapp term, $g_{1/4}$, with the misfit 
which is present even without the anions owing to the proximity to a
quarter filling on one hand, 
and the $g''_{3 \bot}$ term, the combination of both commensurabilities 
of the donors and the anions on the other hand. 

In the present calculation, it is crucial to fix the carrier number 
at the value determined by the anion density.   
However, it is to be noted that if one evaluates the quantity based on eq.(2), 
it results in a deviation of the carrier number from the value in the non-interacting
case, $\Delta N_e = (1/L\pi) \int \d x \langle \partial_x \theta_\rho
\rangle$, as  
\begin{eqnarray}
\Delta N_e / L &=& \frac{2 K_\rho G_{3 \bot}^{''2}}{\pi \alpha} \int_{\alpha}^\infty 
 \frac{ \d r}{\alpha} \left( \frac{r}{\alpha}\right)^{2 - 4 K_\rho} J_1 (q_0 r) \nonumber \\
& & \hspace*{-5em} + \frac{4 K_\rho G_{1/4}^{2}}{\pi \alpha} \int_{\alpha}^\infty 
 \frac{ \d r}{\alpha} \left( \frac{r}{\alpha}\right)^{2 - 16 K_\rho} J_1 (q_0 r) \neq 0, 
\end{eqnarray}
where $G''_{3 \bot} = g''_{3 \bot}/(\pi v_\rho)$, 
$G_{1/4} = g_{1/4}/(2 \pi v_\rho)$, and $J_n(x)$ is the the Bessel function of the
first kind. 
The origin of the deviation is the existence of the misfit parameter.~\cite{Mori-Fukuyama-Imada} 
Therefore, we must add the term, $- (\mu/\pi) \int \d x
\partial_x \theta_\rho$ and keep $\Delta N_e$ to zero. 
To the lowest order of $G''_{3 \bot}$ and $G_{1/4}$, 
the chemical potential $\mu$ is given as,  
\begin{eqnarray}
q_\mu \al 
&=&  - 4 K_\rho G_{3 \bot}''^2 \int_\al^\infty \frac{\d r}{\al} (\frac{r}{\al})^{2 - 4 K_\rho} J_1 (q_0 \al) \nonumber \\
&&- 8 K_\rho G_{1/4}^2 \int_\al^\infty \frac{\d r}{\al} (\frac{r}{\al})^{2 - 16 K_\rho} J_1 (q_0 \al),   
\end{eqnarray}
where $q_\mu = 4 K_\rho \mu / v_\rho$. 
 
To determine the low energy behavior of this effective Hamiltonian, 
we derive the renormalization group (RG) equations by rewriting  the
action $S_\rho$ corresponding to the Hamiltonian, 
${\cal H}_{\rho} - (\mu/\pi) \int \d x \partial_x \theta_\rho$, as 
\begin{eqnarray}
 S_\rho &=& \frac{1}{4 \pi K_\rho} \int \d^2 \vr 
\left\{ (\delx \ti{\theta}_\rho)^2 + (\dely \ti{\theta}_\rho)^2 \right\} \nonumber \\
&+& \frac{G_{3 \bot}}{\pi \al^2} \int \d^2 \vr \cos \left\{ 2 \ti{\theta}_\rho + 3 \chi - (q_{1/4} - q_3'')x \right\}  \nonumber \\
&+& \frac{G''_{3 \bot}}{\pi \al^2} \int \d^2 \vr \cos \left\{ 2 \ti{\theta}_\rho + 5 \chi - q_3''x \right\}  \nonumber \\
&+& \frac{G_{1/4}}{\pi \al^2} \int \d^2 \vr \cos \left\{ 4 \ti{\theta}_\rho + 8 \chi - q_{1/4}x \right\},  
\end{eqnarray}  
where $\ti{\theta}_\rho = \theta_\rho - q_\mu x/2$, $G_{3 \bot}=g_{3
\bot}/(\pi v_\rho)$, $q_{1/4} = q_0 - 2
q_\mu$ and $q''_{3} = q_0 - q_\mu$. 
The condition, $\Delta N_e = 0$, leads to the following self-consistent
equation,   
\begin{eqnarray}
q_\mu \al &=& -4 K_\rho G_{3 \bot}^2 \int_\al^\infty \frac{\d r}{\al} (\frac{r}{\al})^{2 - 4 K_\rho} \ti{J}_1 ((q_{1/4}-q_3'')\al) \nonumber \\
&&- 4 K_\rho G_{3 \bot}''^2 \int_\al^\infty \frac{\d r}{\al} (\frac{r}{\al})^{2 - 4 K_\rho} \ti{J}_1 (q_3''\al) \nonumber \\
&&- 8 K_\rho G_{1/4}^2 \int_\al^\infty \frac{\d r}{\al} (\frac{r}{\al})^{2 - 16 K_\rho} \ti{J}_1 (q_{1/4}\al),  
\end{eqnarray}
where $\ti{J}_1 (x) = {\rm sgn}(x) J_1 (|x|)$.
Eqs.(16) and (17) lead to the following RG equations, 
\begin{eqnarray}
\dl K_\rho &=& - 8 K_\rho^2 G_{1/4}^2 J_0(|q_{1/4} \alpha|) \nonumber \\ 
&&- 2 K_\rho^2 G_{3\bot}^2 J_0(|q_{1/4} \alpha - q_3'' \alpha|) \nonumber \\
&&- 2 K_\rho^2 G_{3\bot}''^2 J_0(|q_3'' \alpha|),  \\
\dl q_{1/4} \al &=& q_{1/4} \al - 16 K_\rho G_{1/4}^2 \ti{J}_1 (q_{1/4} \al) \nonumber \\
&& - 8 K_\rho G_{3 \bot}^2 \ti{J}_1 (q_{1/4} \alpha - q_3'' \alpha) \nonumber \\ 
&& - 8 K_\rho G_{3 \bot}''^2 \ti{J}_1 (q_{3}'' \alpha), \\
\dl q_{3}'' \al &=& q_{3}'' \al - 8 K_\rho G_{1/4}^2 \ti{J}_1 (q_{1/4} \al) \nonumber \\
&& - 4 K_\rho G_{3 \bot}^2 \ti{J}_1 (q_{1/4} \alpha - q_3'' \alpha) \nonumber \\
&& - 4 K_\rho G_{3 \bot}''^2 \ti{J}_1 (q_{3}'' \alpha), \\
\dl G_{3 \bot} &=& (2 - 2 K_\rho) G_{3 \bot} \nonumber \\ 
&& - G_{3 \bot}'' G_{1/4} J_0 (|(q_3'' \al + q_{1/4} \al)/2|), \\ 
\dl G_{3 \bot}'' &=& (2 - 2 K_\rho) G_{3 \bot}'' \nonumber \\ &&- G_{3 \bot} G_{1/4} J_0 (|q_{1/4} \al - q_3'' \al/2 |), \\
\dl G_{1/4}  &=& (2 - 8 K_\rho) G_{1/4} \nonumber \\ && - G_{3 \bot} G_{3 \bot}'' J_0 (|q_3'' \al - q_{1/4} \al/2 |), \\
\dl q_\mu \al  &=& q_\mu \al + 4 K_\rho G_{3 \bot}^2 \ti{J}_1 ((q_{1/4} - q_3'') \al) \nonumber \\
&&+ 4 K_\rho G_{3 \bot}''^2 \ti{J}_1 ( q''_3 \al) \nonumber \\ 
&&+ 8 K_\rho G_{1/4}^2 \ti{J}_1 ( q_{1/4} \al).    
\end{eqnarray}
Eqs.(18)-(23) are obtained for the condition of the action, eq.(16), 
being invariant under RG transformation, whereas eq.(24) is obtained from 
the condition of the chemical potential, eq.(17). 
From eqs.(19), (20) and (24), it is shown that 
the quantity $q_0$, whose dimension is $({\rm length})^{-1}$, 
is scaled as $q_0(l) = q_0 \e^l$. 
This shows the fact that the carrier number is indeed conserved 
 without any effects from the interaction. 

Typical flows of the RG equations are shown in Fig.2 
for $U/t=3.0$ and $V/t=0.0$ where 
the carrier number is fixed as $n=0.441$
taken from the actual material (MDT-TS)(AuI$_2$)$_{0.441}$.  
\begin{figure}
\centerline{\includegraphics[width=7.0truecm]{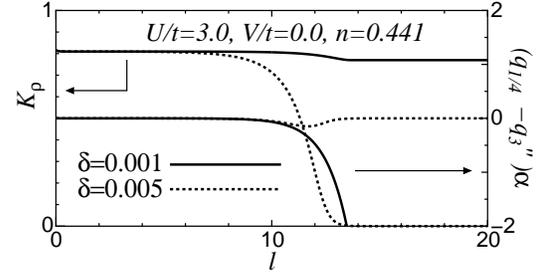}}
\vspace*{-1em}
\caption{
The solutions of the RG equations, $K_\rho$ and $(q_{1/4} - q_3'') \al $
 ( the misfit parameter of the $G_{3 \bot}$-term ) for $U/t=3.0$, $V/t=0.0$ and
 $n=0.441$. 
The cases of $\delta = 0.001$ and $0.005$ are denoted by the solid and
 dotted curves, respectively. 
}
\label{fig:typica-RG}
\end{figure}
In the case of $\delta = 0.005$, 
$K_\rho$ tends to zero implying that the ground state is an insulator, 
due to the commensurability in the half-filling, $G_{3\bot}$. 
This can be seen in the RG equations since 
the misfit parameter in the 
$G_{3\bot}$-term, $(q_{1/4} - q_3'') \al = - q_\mu \al$ vanishes (see Fig.2) 
and then $G_{3\bot}$ affects the renormalization of $K_\rho$ through eq.(18) 
while those in the $G_{1/4}$- and $G_{3 \bot}''$-terms, $q_{1/4}\al$ and $q_3''\al$,  
tend to $\infty$ and then these effects become negligible 
due to the oscillating behavior of the Bessel function. 
Hence the origin of this insulating state is nothing but the commensurate
potential of the effective half-filling generated by the anion potential. 
Namely, the insulating state is indeed the IC Mott insulator.

On the other hand, in the case of smaller potential due to  the anions, $\delta = 0.001$, 
a metallic state with finite value of $K_\rho$ is realized.  
Here in eq.(18) the effects of the $G_{3 \bot}$-, $G''_{3 \bot}$- and $G_{1/4}$-terms 
on $K_\rho$ disappear at the low energy 
since all the misfit parameters are divergent. 
This metallic state is {\it not} realized if we set $G_{1/4}$
to zero.  
Therefore we can state that the origin of the MI transition is 
the interplay between the different kinds of commensurabilities. 

Next, we show ground state phase diagrams as a 
function of the model parameters. 
First,  
the phase diagram  on the plane of $U/t$ and $\delta$ in the case of
$V/t=0$ is shown in Fig.3.
\begin{figure}
\centerline{\includegraphics[width=6.5truecm]{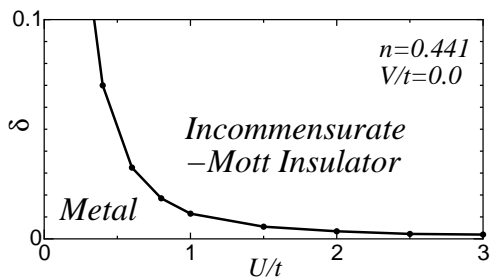}}
\vspace*{-1.2em}
\caption{
The phase diagram on the plane of $\delta$ and $U/t$ in the case of $V/t=0$
 and $n = 0.441$. 
}
\label{fig:phase-V0}
\end{figure}
\begin{figure}
\centerline{\includegraphics[width=6.7truecm]{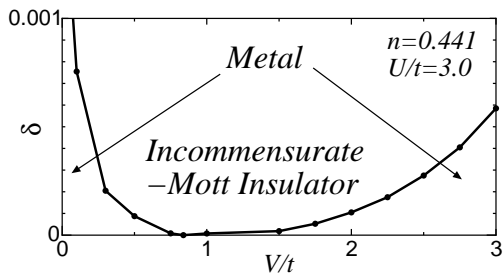}}
\vspace*{-1.2em}
\caption{
The phase diagram on the plane of $\delta$ and $V/t$ in the case of $U/t=3.0$
 and $n = 0.441$. 
}
\label{fig:phase-U3}
\end{figure}
Since the quantity $\delta$ is proportional to $v_0/t$, 
the transition from the metallic state to the IC Mott insulator occurs 
when the potential from the anions increases and/or the band width
decreases. 
When $U$~$\to$~$\infty$, the present system can be mapped onto 
a non-interacting spinless Fermion system with the Fermi wavenumber doubled, as $2\kf$.\cite{Ovchinnikov} 
In this case the insulating state is
realized by an infinitesimal $\delta$ 
because a gap opens at $\pm 2\kf$ (see Fig.1), consistent with Fig.3. 

The role of the $G_{1/4}$-term on the MI transition 
becomes clearer when $V$ is varied. 
It is because the coupling constant $G_{1/4}$ changes its sign when $V$ increases.\cite{Yoshioka}  
We show the phase diagram on the $V/t$ - $\delta$ plane in Fig.~\ref{fig:phase-U3} 
for $U/t = 3.0$. At $V=V_c=0.838 t$ where $G_{1/4} = 0$, 
the IC Mott insulating state is realized for infinitesimal $\delta$. 
For $V >V_c$, the absolute value of $G_{1/4}$ increases again
and results in a finite metallic region. 
Therefore, a re-entrant transition, metal $\to$ IC Mott insulator $\to$ metal, 
occurs when $V$ increases. 
Note that there is no qualitative difference between the metallic states 
in the two distinct regions.    

Finally let us discuss the relevance of our results to the experiments. 
The difference of the ground state in metallic MDT-TSF and 
MDT-ST compounds and that in the MDT-TS compound undergoing MI crossover 
can be naturally understood as follows. 
The extended H$\ddot{\rm u}$ckel scheme 
provides transfer integrals, \ie the band width, of the MDT-TSF families 
larger than that of the MDT-TS compound~\cite{Kawamoto,Kawamoto-5}, 
which is consistent with our results that the decrease of 
the bandwidth lead to an MI transition, as seen in Fig.3. 
In our 1D model the spin degree of freedom is essentially that of 
the 1D Heisenberg model showing no magnetic order. 
However, in the IC Mott insulating state in the actual 2D material 
we generally expect that antiferromagnetic order appears 
at low temperature due to the three dimensionality, 
as in fact observed.~\cite{Kawamoto-5} 
In this case, the magnetic ordered moment should be large, 
compared to, e.g., that of
the spin-density-wave state due to the nesting of the Fermi surface. 

In conclusion, we investigated 
the electronic state of the one-dimensional extended Hubbard model 
close to quarter-filling under an incommensurate anion potential. 
We found that a transition between the metallic state and 
an incommensurate Mott insulator can occur, 
whose origin is the interplay between 
the commensurability energy generated by the anion potential and 
that in the donor lattice. 
To the authors' best knowledge this is the first theoretical study of a ``Mott transition" 
generated by such interplay between different commensurabilities. 
It would be interesting to investigate the critical properties of this transition 
in the actual compounds and compare with the ``usual" Mott transition seen in 
the typical 2D $A_2B$ molecular conductors, $\kappa$-(BEDT-TTF)$_2X$,  
which is recently attracting interests.~\cite{Kagawa}

The authors would like to thank T. Kawamoto for sending them his
preprint prior to publication. They also acknowledge 
G. Baskaran, M. Ogata, K. Kanoda, and J. Kishine for valuable discussions and comments.
This work was supported by Grant-in-Aid for Scientific Research on
Priority Area of Molecular Conductors (No.15073213) and 
Grant-in-Aid  for Scientific Research (C) (No. 14540302 and 15540343)
from MEXT.

\end{document}